# CONSTRAINTS ON THE OPACITY OF SPIRAL DISKS FROM NEAR-INFRARED OBSERVATIONS

HANS-WALTER RIX
*Institute for Advanced Study, Princeton, NJ 08540, USA*
and
*Max-Planck-Institut für Astrophysik, 85740 Garching, Germany*

## 1. Introduction

In this paper I review how near-infrared (NIR) observations can constrain the opacity of spiral disks. Basic considerations show that NIR photometry provides a powerful probe of the optical depths in spiral galaxy disks in the regime of interest $\tau_V \sim 1$. I review the existing opacity constraints from the analysis of dust lanes in edge-on and face-on galaxies. The "internal extinction correction" in the NIR-Tully-Fisher relation deserves particular attention as the most powerful constraint on the impact of dust on the total luminosity of spiral galaxies. All observations for luminous spirals point towards an effective, face-on optical depth of $\langle \tau_V \rangle = 0.5 - 1$.

## 2. Which Near-Infrared: $0.8\mu$m or $2.2\mu$m?

The "near-infrared" wavelength region spans a factor of three in wavelength, $0.8\mu m < \lambda < 2.4\mu m$. This region encompasses a wide range of difficulties in the data taking and a comparably wide range in potential for the data interpretation. Basically, data beyond $1\mu$m are much harder to acquire, but the greatly reduced dust opacities at larger wavelength often provide a crucial advantage in the data interpretation.

### 2.1. GOING TO $2.2\mu M$: THE PAIN

An overview of the differences in the data acquisition between I($0.8\mu$m) and K($2.2\mu$m) is given in Table 1. The first striking difference between CCDs and the HgTeCd arrays (such as the NICMOS3) is their size: currently available CCD have almost 100 times more pixels! This becomes a decisive advantage when imaging nearby galaxies, where the angular size of the target is larger than the field-of-view of the detector and mosaicing may be necessary. Both types of detectors have comparable quantum efficiency, and for both detector types does the photon noise dominate all sources of detector noise. The second large difference between the observational set-ups is independent of the detectors: at one exponential disk scale length, the galaxy surface brightness at $0.8\mu m$ is still 1/2 that of the sky. For the K-band, this ratio is only 0.005, the sky signal dominates everywhere. Fortunately, this large background signal can be used very effectively to "flat-field" the image. Sensitivity variations among the pixels are routinely corrected to a few parts in $10^4$.

The bottom line of all these considerations is given at the end of Table 1: for a given amount of observing time, the S/N achievable at $0.8\mu m$ is ten times higher than at $2.2\mu m$. Put differently, to achieve a comparable signal-to-noise one would need to observe 100 times longer in the K-band.

TABLE 1. Detector Comparison for Near-IR Imaging

| Comparison | CCD($0.8\mu m$) | NICMOS3($2.2\mu m$) | Ratio |
|---|---|---|---|
| Nr. of Pixels | $2048 \times 2048$ | $256 \times 256$ | 65 |
| Quantum Efficiency | $\gtrsim 40\%$ | $\gtrsim 50\%$ | $\sim 1$ |
| Detector Noise | negligible | negligible | $\sim 1$ |
| Flat-Fielding | $\gtrsim 10^{-3}$ | $\gtrsim 10^{-4}$ | 0.1 |
| Gal. / Sky Contrast[1] | 0.5 | 0.005 | 100 |
| S/N (asec$^{-2}$min$^{-1}$)[1] | 57 | 5 | 11 |

[1] Measured at $R \sim R_{exp}$

## 2.2. GOING TO $2.2\mu M$: THE GAIN

The above comparison raises necessarily the question why one should bother taking data beyond $1\mu m$. The answer is illustrated in Figure 1. While the dust opacities and albedos in the I band are still very similar to the other "optical" wavelength bands, they are dramatically different at K($2.2\mu m$). The opacity data in Figure 1 are taken from Laor and Draine (1993 and *pers. comm.*); note that the albedo only refers to the isotropic part of the scattering cross section. Most theoretical calculations, and many empirical determinations, show that in spiral disks the dust optical depth in I($0.8\mu m$) is six times higher than in K($2.2\mu m$). For the interpretation of data this difference has several consequences: (1) A lower albedo makes the interpretation of data by radiative transfer models easier. (2) Combining optical and NIR data provides a long wavelength baseline for color and reddening analyses. (3) In many cases (see contributions throughout these proceedings) the dust extinction is negligible only at K($2.2\mu$m), but not yet at I($0.8\mu$m). Therefore, data at $2\mu$m can often provide an empirical template for a galaxy's appearance without dust. (4) In dust lanes and in edge-on galaxies the optically thin regime ($\tau_\lambda < 1$) is only probed at wavelengths as long as $\lambda \sim 2\mu$m.

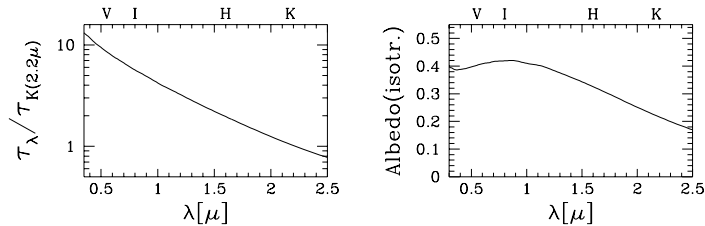

Fig.1: The left panel shows the dramatic drop in the dust absorption cross-section from $0.8\mu$m to $2.2\mu$m, taken from Laor and Draine (1993). The right panel shows the albedo, to isotropic scattering, for the same wavelength range; it peaks near the I-band.

## 3. For What Dust Signatures Should One Look?

### 3.1. OBSERVATIONAL SIGNATURES

Most observational test for the presence of dust, using optical-IR data, fall into three categories:

• Study the surface brightness in the presence of dust, $\mu(\vec{R}, \tau_\lambda)$ in highly inclined galaxies, if the dust-free surface brightness, $\mu(\vec{R}, \tau_\lambda = 0)$, is known independently. Most commonly this independent information is obtained through the (assumed) point-symmetry of the dust-free galaxy.

• Study the surface brightness $\mu(\vec{R}, \tau_\lambda)$, but obtain $\mu(\vec{R}, \tau_\lambda = 0)$ directly from NIR observations, if dust extinction can be neglected there. This test, applicable e.g. to dust lanes in face-on galaxies, requires the assumption that the colors of the dust-free galaxy can be determined empirically, e.g. from patches next to the dust lanes.

• Study the surface brightness, total luminosity or color of intrinsically identical galaxies as a function of their inclination $\cos i$. These tests exploit the disk geometry of the dust distribution, which results in increasing extinction towards edge-on orientations. A suitable, dust-independent, property to find identical galaxies is, e.g., their HI rotation speed (see Section 4.3).

### 3.2. A RADIATIVE TRANSFER MODEL

As discussed throughout this meeting, the conversion of any observations into a statement about the optical depth, $\tau_\lambda$, is impossible without a radiative transfer model. To relate the observational signatures quantitatively to optical depth estimates, we use here a model, described by Rix and Rieke (1993). [For other radiative transfer models, see Elmegreen (1980), Kylafis and Bahcall (1987) and Witt *et al.* (1992).] Briefly, this model assumes plane-parallel geometry, a Gaussian distribution of dust and stars in the vertical direction (with scale heights of $H_D$ and $H_S$, respectively) and a homogeneous distribution of dust with a *face-on* optical depth of $\tau_V$. This model configuration can be viewed from any angle, $i$, except exactly edge-on ($i = 90°$).

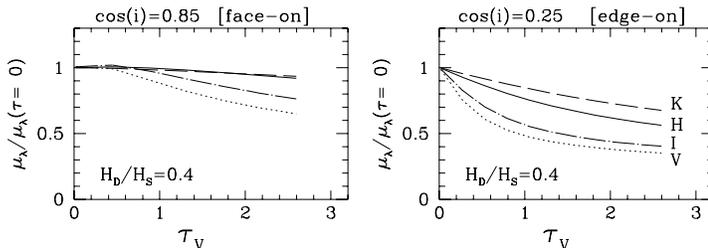

Fig 2.: Decrease in surface brightness (or luminosity) with increasing optical depth, $\tau_V$, predicted by the radiative transfer models for the V, I, H and K band. The left panel shows the case of a nearly face-on galaxy, the right panel a nearly highly inclined galaxy.

Dust scattering is treated in the single scattering approximation and the absorption and scattering properties of the dust are taken from Laor and Draine (1993, and *pers. comm.*). Each model is completely specified by the three parameters: $H_D/H_S$, $\tau_V$ and $\cos i$. The model predictions are applicable in two regimes: (a) when making a local comparison of surface brightnesses from a nearly face-on direction (e.g. when analyzing a dust lane in a face-on galaxy), and (b) when estimating changes in the global luminosity or color due to dust. The model provides a good estimate of the "effective" optical depth (i.e. the optical depth of a homogeneous dust distribution causing the same observational effects), as long as the dust is spatially more extended than the luminosity sources (stars).

Figures 2 and 3 illustrate the model expectation for the observational tests in Section 3.1. Figure 2 shows how the surface brightness (or total luminosity) changes in various pass-bands as the optical depth increases. Figure 3 shows how the observable surface brightness deviates from the dust-free case as the inclination increases. Note that in both figures the I band (dashed-dotted line) is much closer in its behaviour to the V band (dotted line) than the K band (dashed line).

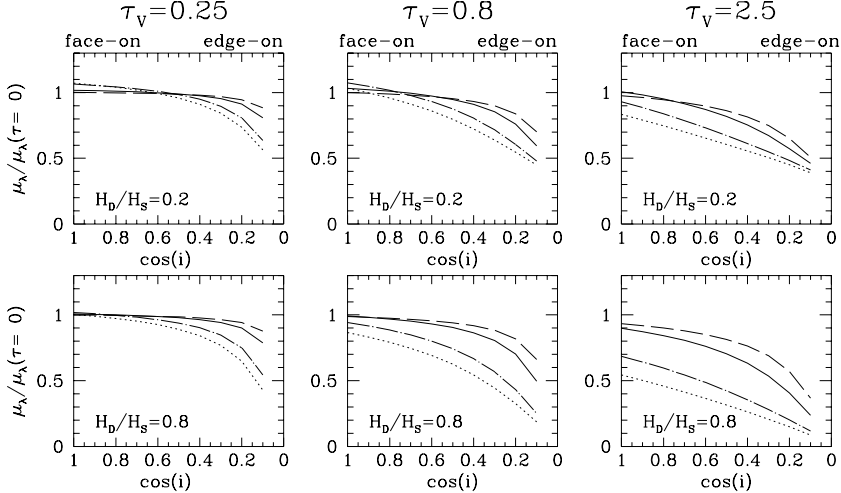

Fig 3.: Decrease in surface brightness (or luminosity) with increasing inclination for three different values of the (face-on) optical depth, $\tau_V$, and for two different scale heights of the dust. The four different lines refer to the V, I, H and K band, as in Figure 2.

## 4. Observational Signatures

### 4.1. DUST LANES IN EDGE-ON GALAXIES

Optical and NIR photometry of dust lanes in nearly edge-on galaxies provides one of the most straightforward probes of the dust opacity in disks (Knapen *et al.* 1991, Barnaby and Thronson, 1992, Byun, 1993, Jansen *et al.*, 1994). Such opacity tests rely only on the reflection-symmetric distribution of the stellar luminosity sources, and on the assumption that the dust is distributed in a disk co-planar to the stars. In its simplest form the test consists of comparing the brightness at a point on the near side of the minor axis, $\mu_{near} = \mu_{inside} e^{-\tau} + \mu_{outside}$ to the corresponding point on the far side of the minor axis $\mu_{far} = \mu_{outside} e^{-\tau} + \mu_{inside}$. Here $\mu_{near}$ (or $\mu_{far}$) is the surface brightness arising from radii smaller (or larger) than the dust annulus under consideration. If scattering is neglected and if $\mu_{inside} \gg \mu_{outside}$ holds (e.g. because all the bulge luminosity is inside the dust seen in projection), then the optical depth can be obtained immediately by ratioing $\mu_{near}$ and $\mu_{far}$. NIR observations are crucial because for a simple, model-independent $\mu_{near}/\mu_{far}$ test one has to assume not only that $\mu_{inside} \gg \mu_{outside}$ but also that $\mu_{inside}e^{-\tau_\lambda} \gg \mu_{outside}$, which is more likely to holt for small $\tau_\lambda$. Most recently, Jansen *et al.* (1994) observed four galaxies with edge-on dust lanes in various bands, ranging from B(0.44$\mu$m) to K(2.2$\mu$m). They find that the optical depths, at V, are $0.2 - 2$ *in projection*. These correspond to face-on optical depths (multiplying by $\cos(i)$), of only a few tenths. However, three of their four sample galaxies are early type spirals and these conclusions cannot be generalized to other types. Given the galaxy's inclination, the run of $\tau_\lambda$ along the projected minor axis can also give the true radial profile of the optical depths: the dust extends typically to 3 (stellar) scale-lengths (e.g. Knapen *et al.* 1991) but with a radial profile less steep than the stars'.

## 4.2. DUST LANES IN FACE-ON GALAXIES

In this Section I propose a new type of test: rather than taking the dust free case as the reference observation, one could take the completely opaque case. This way, one can constrain *locally* the optical depth in face-on disks at any point next to a dark dust lane. The constraint comes from a quantitative answer to the question "How much dust can there be in the rest of the disk before the contrast (in intensity or color) to the dust lane is spoiled?".

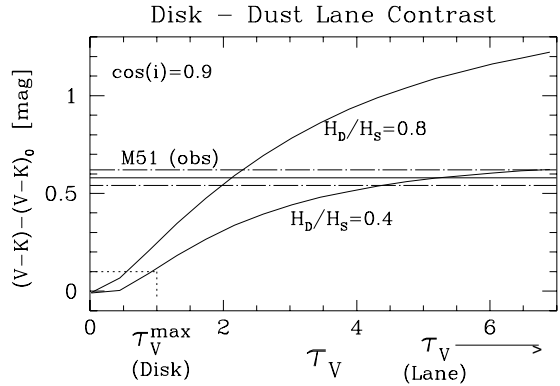

Fig. 4: V-K color contrast between a point in the disk [with $\tau_V$(Disk)] and a dark dust lane [with $\tau_V$(Lane)], predicted by models with $H_D/H_S = 0.4$ and 0.8. The dust free color of the galaxy is denoted by $(V-K)_0$. The observed dust lane - disk color contrast in M51 (RR93) of $0.58 \pm 0.04$ is well matched by $\tau_V$(Lane) $\geq 5$ and $\tau_V$(Disk) $\ll 1$. The observed width of the dust lane $< 100$pc provides an independent constraint on its scale height $H_D/H_S \lesssim 0.4$. If the disk had an optical depth $\tau_V^{max}$(Disk) $\gtrsim 1$ throughout (i.e. $(V-K) - (V-K)_0 > 0.1$ for $H_D^{Disk}/H_S = 0.4$), no amount of dust in the lane could lead to the observed disk-lane color difference. If the "diffuse" dust throughout the disk has a scale height higher than the lane's, then the limit on $\tau_V^{max}$(Disk) tightens.

It is further necessary to assume that the color of the stellar population in and next to the dust lane changes only little. If suitable color bands (e.g. R and K) are chosen, and if regions of intense star formation are avoided, this should be an acceptable assumption (see e.g. RR93). The test is strengthened if the "diffuse" dust throughout the disk is assumed to be no flatter than the dust lane. As a sample application, we select one of the dust lanes ("Lane 45") in M51 from RR93. The projected width of this dust lane (in the disk plane) is only 80pc, and we assume that its vertical scale height is not much larger than that. Given the independent information (from other galaxies, e.g. Elmegreen, 1980) on the scale heights of the V and K band stellar light, this leads to $H_D^{lane}/H_S \lesssim 0.4$ and $H_D^{lane} < H_D^{diff.}$. Figure 3 now shows that (using the models from Section 3) a disk − dust lane color difference of $V - K = 0.58$ is only possible as long as the optical depth outside the lane is less than $\tau_V \lesssim 1$. Even though this technique deserves more detailed modelling than presented here, it clearly is a way to rule out uniformly large optical depths in the inner portions of spiral disks whenever prominent dust lanes are visible.

TABLE 2. Comparison of "Internal Extinction" Corrections

| Authors | Wavelength | Relation | Fitted Param. | Remarks |
|---|---|---|---|---|
| RC3 | B(0.45$\mu$) | $A_i = \alpha \log(\frac{a}{b})$ | $\alpha = 1.45$ | type dependence |
| PW 92 | H(1.6$\mu$) | $\mu = \mu_0 + 2.5\, c\, \log(\frac{a}{b})$ | $c = 0.8$ | with apert. corr. |
| Han 92 | I(0.8$\mu$) | $\mu = \mu_0 + \tilde{c} \log(\frac{a}{b})$ | $c = 1.9$ | $R \geq 3 R_{exp}$ |
| B 94 | I(0.8$\mu$) H(1.6$\mu$) | $m = m_I(0) + a_I \left(1 - \frac{b}{a}\right)$ | $a_I = 1.4$ $a_H = 0.3 \pm 0.2$ | 23 galaxies |
| G 94 | I(0.8$\mu$) | $m = m_I(0) + \gamma \log(\frac{a}{b})$ | $\gamma = 1.15 \pm 0.2$ | |

RC3: de Vaucouleurs et al. 1991; PW 92: Peletier and Willner, 1992;
B 94: Bernstein et al. 1994; G 94: Giovanelli et al. 1994

4.3. THE "INTRINSIC EXTINCTION" CORRECTION FOR THE NEAR-IR TULLY FISHER RELATION

The tight relation between the stellar luminosity of a galaxy and its rotation velocity (known as the "Tully-Fisher-Relation", hereafter, TF relation), provides an excellent test for the impact of dust on a galaxy's apparent *total* magnitude. If spiral galaxies were dust-free, there should be no correlation between the apparent magnitudes, $m(i)|_{W_{20}}$, and the inclinations, $i$, for galaxies with identical rotation speed $W_{20}/\sin(i)$ (as measured through their HI linewidth $W_{20}$). However, such a correlation, in the sense that more edge-on galaxies appear fainter, has been known from the first studies using B-band photometry. If this B-band "internal extinction correction" (e.g. from RC3) were to scale to the near-IR as the standard extinction curve, it should be negligible there. However, all recent TF work in the NIR has found this correction to be significant (see Table 2).

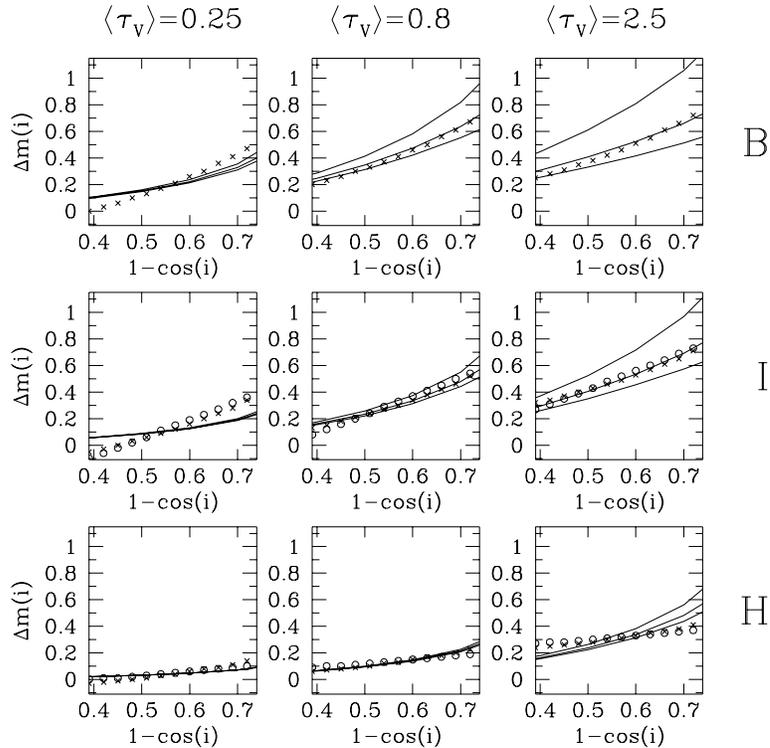

Fig. 5: Comparison of the observed "internal extinction corrections" (points) with radiative transfer models (solid lines). Each panel shows the increasing magnitude difference, $\Delta m(i)$, between a dust free galaxy and a dusty galaxy, as the galaxies become more edge-on (increasing $1-\cos i$). The panels show $\Delta m(i)$ for three different optical depths and for three different wavelengths. The three lines in each panel represent (top to bottom) models with $H_D/H_S = 0.8, 0.4, 0.2$. The circles represent the I and H band extinction corrections from Bernstein et al. (1994); the crosses are the fits to the data from RC3 in B, Giovanelli et al. (1994) in I and Peletier and Willner (1992) in H. Note that the any vertical offset between the models and the data is arbitrary, because the dust-free luminosity is not an observable. Models with an effective optical depth of $\langle \tau_V \rangle \sim 0.8$ match the data well in all three wavelength bands.

In practice, this correction is determined as follows. Choose a parametrized functional form for the change of apparent brightness with inclination for intrinsically identical galaxies; all authors have used *ad hoc* forms, which are listed in Table 2. Then determine the parameters that minimize the scatter in the TF relation [now $m \equiv m(i, W_{20}/\sin i)$] from the data, usually over the range $0.2 \lesssim b/a \lesssim 0.6$. The results are listed in Table 2. Even though various authors use different parametrizations and differing samples, the resulting "internal extinction corrections", $m(i)|_{W_{20}}$, are mutually consistent, both in the I-band (Bernstein *et al.*, 1994; Giovanelli *et al.*, 1994) and the H-band (Peletier and Willner, 1992; Bernstein *et al.*, 1994). This is illustrated by the circles and crosses in Figure 5.

Using the models from Section 3.2, we can estimate the luminosity-weighted, effective, face-on optical depth, $\langle \tau_V \rangle$, of the galaxies. We consider models with three different optical depths ($\langle \tau_V \rangle = 0.25, 0.8, 2.5$) and with three different scale height ratios ($H_D/H_S = 0.2, 0.4, 0.8$). The predictions for all these models are compared to the observed relations in Figure 5. While the data cannot distinguish between models with different scale height ratios, they can discriminate between models of different optical depths. The model with $\langle \tau_V \rangle = 0.25$ clearly under-predicts the slope of the $m(i)|_{W_{20}}$ vs $\cos i$ relation in the I (and B) band and the model with $\langle \tau_V \rangle = 2.5$ clearly over-predicts the slope in the H-band. However, the model with $\langle \tau_V \rangle = 0.8$ matches all bands well.

These results lead to the conclusion that the Sb-Sc galaxies used in the TF studies have an internal dust extinction *equivalent to a uniform, homogeneous dust layer of* $\tau_V \sim 0.8$.

### 4.4. THE INCLINATION DEPENDENCE OF COLORS AND COLOR GRADIENTS

Rather than considering only the total luminosity as a function of inclination, one could design stronger tests by considering simultaneously the change in luminosity, optical-NIR color and color gradient as a function of inclination. The color gradients can probe the radial profile of effective optical depth, and thus provide a new piece of information. Even though optical-NIR photometry for samples of highly inclined (Terndrup *et al.* 1994) and face-on (de Jong and van der Kruit, 1994) galaxies have been published, no consistent, large data set, sufficient to test $\partial(\text{color})/\partial(\cos i)$ and $\partial(\text{color grad.})/\partial(\cos i)$ exists to date. However, such data sets are currently being assembled.

## 5. Conclusions

The purpose of this talk has been two-fold. First it tried to give some practical guidelines for designing (and interpreting) observational opacity tests using NIR observations. Second, it tried to present some of the opacity constraints obtained to date.

Several points are worth stressing again in conclusion:

• NIR data beyond $1\mu m$ [especially at $K(2.2\mu m)$] are much more difficult to obtain, but may provide clues that cannot be obtained from data at shorter wavelengths. Even within the NIR the choice of the wavelength region merits thorough consideration.

• The study of a small number of nearly edge-on galaxies with dust lanes shows that in these objects the optical depths (along the line of sight) range from $0.2 \lesssim \tau_V \lesssim 2$.

• The existence of a large color and intensity contrast between the prominent dust lanes (e.g. in M51) and the rest of the disk shows that even the inner portions of disks must have $\tau_V \lesssim 1$.

• The "internal extinction corrections" in the Tully-Fisher relations measured by various authors and in various bands (B, I, H) are *mutually consistent*. All these corrections are matched well by a simple radiative transfer model assuming a mean, face-on optical depths of $\langle \tau_V \rangle \sim 0.8$.

Each of these tests is in need of further modelling. A proper radiative transfer model should produce a robust estimate of the radial profile of $\tau_V$ from the nearly edge-on galaxies. Modelling the disk - dust lane color contrast more carfully will provide *localized* measurements of $\tau_V$ in the inner disk portions. Finally, it should be checked whether these local constraints are consistent with the global ones derived from the Tully-Fisher relation.